\def\fnote#1{\footnote}

\def\cwleftpar#1#2{\leftskip #1 \rightskip #2 plus 1fill}
\def\cwrightpar#1#2{\leftskip #1 plus 1fill \rightskip #2}
\def\cwcenterpar#1#2{\leftskip #1 plus 1fill \rightskip #2 plus 1fill}
\def\cwfullpar#1#2{\leftskip#1\rightskip#2}

\def\cwoutdent#1#2{\llap{\hbox to #1{#2 \hss}}\ignorespaces}
\def\cwparbegin#1#2#3#4#5{
	\ifcase #1 \cwleftpar{#2}{#3}
	\or \cwrightpar{#2}{#3}
	\or \cwcenterpar{#2}{#3}
	\else \cwfullpar{#2}{#3}\fi
	\ifcase #4 \baselineskip = 1.5\baselineskip
	\or \baselineskip = 2\baselineskip
	\or \baselineskip = 3\baselineskip
	\else \baselineskip = 1\baselineskip\fi
	\ifdim #5 > 0in \else \noindent \fi
	\noindent\ignorespaces}
\documentstyle[12pt]{article}
\begin{document}
\input amssym.def

\begin{flushright}
\mbox{\large\bf\vbox{
\hbox{Preprint CLNS 945/89}\hbox{Cornell University,}\hbox{USA, 1989.}}}  
\end{flushright}

\begin{center}
{\bf Flag Spaces in KP Theory and Virasoro Action on  
$\det D_j$ and Segal-Wilson $\tau$-Function}\footnote 
{\bf A slightly modified
version of this text was published in ``Research Reports in Physics. Problems
of Modern Quantum Field Theory.'' Editors: A.A.Belavin, A.U.Klimyk, 
A.B.Zamolodchikov. Springer-Verlag Berlin, Heidelberg 1989, pp. 86--106.}
\end{center}

\medskip
\begin{center} {\large P.G.Grinevich}$^{\dag}${\large , A.Yu.Orlov}$^{\ddag} $
\end{center}
\par
\noindent $^{\dag}$Landau Institute for Theoretical Physics, Kosygina 2,
Moscow, 117940, USSR
\par
\noindent $^{\ddag}$Oceanology Institute, Krasikova 23, 117218, Moscow, USSR.
\par
\medskip
\begin{center} {\bf Abstract.}

We consider Virasoro action on flag spaces corresponding to
the Riemann surfaces with two marked points.
\end{center}
\par
\medskip
{\bf 0. Introduction.} The development of string theory and conformal
theories on Riemann surfaces has produced interest in the objects of
soliton theory (see, for example, \cite{1}). There are a number of papers
using the Segal-Wilson Grassmannians as a model of universal moduli
space -- the space containing all the Riemann surfaces of finite
genus. In the case of superstrings, which appears to be simpler, the
measure was calculated in \cite{2}. Another soliton object -- the 
$\tau$-function introduced by the Kyoto mathematicians (see \cite{28},
\cite{3} and references therein) -- may be defined as some vacuum 
expectation of fermionic fields \cite{4}. The monodromy properties of
the $\tau $-function \cite{5} were used to calculate the 
$\det \bar{\partial }$ for hyperelliptic curves \cite{6}.
\par
The Segal-Wilson Grassmannians correspond to Riemann surfaces
with a single marked point. However it is more natural to consider
Riemann surfaces with a number of marked points. In the simplest case
we have two marked points which correspond to in and out states of
the string. Conformal field theory on such surfaces was constructed
by I.M.Krichever and S.P.Novikov \cite{7}-\cite{9}, where they introduced 
some analogue of the Laurent basis for tensor fields on such surfaces. We
show that the proper analogue of the Grassmannian in this situation
is the flag space.
\par
In conformal theories the following algebras play a crucial
role --  the algebra of the vector fields on the circle and its
central extension known as the Virasoro algebra. In the Krichever-Novikov
basis the Virasoro algebra appears to be generalized-graded.
\par
In our paper we consider the following object: a Riemann surface 
$\Gamma $, with two marked points $P_{0}$ and $\infty $ and divisor 
$\gamma _{1},\ldots  ,\gamma _{g}$. The points $P_{0}$ and $\infty$ 
play different roles in our approach. We consider a small contour 
$S$ around $\infty$, and  the space ${\cal L}^{2}(S)$ of a set of 
elements $W(t_{0}), t_{0}\in {\Bbb Z}$, where $W(t_{0})$ consists of 
functions meromorphic outside $\infty$ with divisor 
$t_{0}P_{0}-\gamma _{1}-\ldots  -\gamma _{g}, W(t_{0}+1) \subset  W(t_{0}), 
W=\{W(t_{0})\}$ is an element of
the flag space. If $\gamma _{i}$ are located in the points $P_{0}$ and 
$\infty $, then $W(t_{0})$ are generated by the elements of the 
Krichever-Novikov basis. We also fix a local parameter $z$ in $\infty $. 
The Krichever construction (see review, \cite{16}) allows us to construct 
solutions of the Kadomtsev-Petviashvili equation via such data depending on 
an extra parameter $t_{0}$. In the KP theory, $t_{0}$ was introduced
in \cite{3},\cite{27},
where $t_{0}$ was treated as a discrete time in the generalized hierarchy 
containing KP and Toda lattice hierarchies. In string theory, $t_{0}$ plays 
the role of momentum.
\par
In our paper we study the Virasoro action on the KP theory
objects generated by the Virasoro action on the Riemann surfaces. Our
main tool is the Cauchy-Baker-Akhiezer kernel (see section 2.4),
which inverts the $\bar{\partial}$ operator on a certain bundle, 
${\Bbb B}_{j}(t_{0},\vec{t},D)$.  With
the help of this kernel we present an explicit version of the
Segal-Wilson construction.  We show that the Virasoro action on the
KP potentials coincides with the non-isospectral KP-hierarchy \cite{10}.
The times of equations from this hierarchy correspond to the
deformations of the Riemann surfaces and form the coordinates on the
moduli space. We also calculate the action on the Baker-Akhiezer
function, i.e., solve the problem by I.M.Krichever and S.P.Novikov.
In these cases we have no central extension.  Then we introduce the
$\tau$-function corresponding to the $j$-tensors, and calculate the Virasoro
action on it. This action is represented by second order differential
operators acting on the space of functions of an infinite number of
variables. The central charge is $c_{j}=6j^{2}- 6j + 1$ in accordance with
\cite{11}. (The Virasoro action on the $j$-tensors Grassmanniansand central
extensions were considered in \cite{2}). This representation is valid for
arbitrary $\tau $-functions.  In the case of the algebro-geometrical
$\tau $-function, we have $\theta$-functional realizations of generalized 
Verma modules in the sense of \cite{7}-\cite{9}, parameter $t_{0}$
playing the role of highest weight. "Naive" calculation of the 
variation of the $\det \bar{\partial }_{j}$ in
the corresponding bundle gives the same results as the calculation of
the $\tau $-function variation. Nontrivial bundles are necessary to
suppress $(2j-1)(g-1)$ zero modes of operator $\bar{\partial }_{j}$.  
We use the bundle corresponding to the Baker-Akhiezer functions.  
So we may treat the Segal-Wilson $\tau $-function as 
$\det \bar{\partial }_{j}$.  We discuss the connection between the 
$\tau$-function and the Krichever-Novikov vacuum expectation ${\cal A}$
\cite{8},\cite{9}.
\par
\medskip
\noindent {\bf Chapter 1. Riemann surfaces and bundles.}
\par
\medskip
1.1. {\em $\bar{\partial }_{j}$-operator and index. The Cauchy kernel.} 
Consider the following equation on the Riemann surface $\Gamma$ of genus $g$:
$$
\bar\partial f = \varphi
\eqno(1.1)
$$
in the simplest case, when $f$ is a function and $\varphi$ is a $(0,1)$-tensor
i.e. $\varphi=\tilde\varphi(z,\bar{z})d\bar{z}$.  
The difference between the dimensions of $\hbox{Ker}\ \bar{\partial}$
and $\hbox{Coker}\ \bar{\partial}$ is called the index of 
$\bar{\partial}$. Assume that $\bar{\partial}$ maps the nonsingular functions 
to the nonsingular forms. In this case $\bar{\partial }$ has
one-dimensional kernel $(\bar{\partial}\cdot\hbox{const}=0)$ and a 
$g$-dimensional cokernel -- that is, $\hbox{index}\ \bar{\partial } = 1-g$.  
For (1.1), the $g$-dimensional cokernel means that (1.1) has a solution 
if and only if $\varphi $ satisfies $g$ linear relations:
$$
\int\int_{\Gamma}\tilde \varphi(z,\bar{z}) \tilde w_{n}(z) dz d\bar{z} = 0, 
\ n=1,2,\ldots ,g,
\eqno(1.2)
$$
where $\tilde w_{n}(z) dz$ are the holomorphic differentials (see (1.4)).
\par
In the Quillen theory of $\det\bar\partial$ \cite{25}, index
$\bar{\partial}$ is 
assumed to be zero. We can make $\bar{\partial}$ be of index zero and 
invertible by assuming $f$ and $\varphi $ to be elements of the following 
nontrivial bundles. Let $\gamma _{1},\ldots  ,\gamma _{g},P$ be a collection 
of points of general position.  The functions $f$ and $\varphi$ are smooth 
everywhere, except for singularities of the simple poles type 
$\varphi _{k}(z,\bar{z})/(z-\gamma _{k})$ in $\gamma _{k}$ with smooth 
$\varphi _{k}$. $f$ and $\varphi$ have simple zeroes in $P$, of the form 
$d(z,\bar{z})(z-P)$.  Then (1.1) has a unique solution:
$$
f(z,\bar{z}) = \int \int \tilde \omega(z,z'{}) \tilde \varphi(z'{},\bar{z}'{})
dz' d\bar{z}' ,           
\eqno(1.3)
$$
where the Green function $\omega =\tilde\omega(z,z'{})dz'$ has the 
following properties:
\par
1) $\omega $ is a meromorphic 0-form in $z$ and a meromorphic 1-form in $z'$;
\par
2) $\omega $ has simple poles (zeroes) in $\gamma _{1},\ldots ,\gamma_{g}$ and 
a simple zero (pole) in $P$ as a function of $z$ (of $z'$), respectively;
\par
3) $\omega  \simeq  (2\pi i)^{-1} dz'/(z'{}-z)$ as $z \rightarrow  z'{}$.
\par
$\omega $ is the meromorphic analogue of the Cauchy kernel on the
Riemann surface \cite{13}.
\par
The functions $f$ and $\varphi $ can be interpreted as smooth sections of
nontrivial holomorphic bundles.
\par
{\it Remark}. Even in the simplest case of $g=0$ and the ordinary Cauchy
kernel, bundles are nontrivial and correspond to a simple zero in $\infty$.
\par
\medskip
1.2. {\em Notation.} In our paper we assume that the Riemann surfaces
$\Gamma $ are compact.  Let $g$ be the genus of $\Gamma,\ g<\infty $.  
We will need the following geometrical objects on $\Gamma$
\cite{14},\cite{15},\cite{16}: 
\par
1) $\vec{{\bf w}} = ({\bf w}_{1},\ldots  ,{\bf w}_{g})$ - the basis of 
holomorphic 1-differentials with the following standard normalization
$$
\oint _{a_{i}}{\bf w}_{k}= \delta _{ik}, \oint _{b_{i}}{\bf w}_{k}= B_{ik},
\eqno(1.4)
$$
$B_{ik}$ is called the matrix of periods or Riemann matrix.
\par
2) The Abel transformation.  Let $P$ be a collection of points
$P=(P_{1},\ldots  ,P_{n})$.  Then
$$
\vec A(P) =\int_{\infty}^{P_1}\vec {\bf w} + \int_{\infty}^{P_2}\vec {\bf w}
+\ldots + \int_{\infty}^{P_n}\vec {\bf w}.
\eqno(1.5)
$$
3) The prime-form $E(\gamma,\gamma '), \gamma,\gamma' \in \Gamma$ which is 
a holomorphic $-1/2$-form in $\gamma$ and in $\gamma '$ with the 
following properties:
\par
a) $E(\gamma ,\gamma '{}) = 0$ if and only if $\gamma  = \gamma '{}.$
\par
b) Let $t$ be a local coordinate on $\Gamma $.  Then for $\gamma \rightarrow  
\gamma '{}$:
$$
E(\gamma ,\gamma') = \frac{(t(\gamma )-t(\gamma '))
\{1+O((t(\gamma )-t(\gamma '{}))^{2}\}}
{\sqrt{dt(\gamma) dt(\gamma')}}  .   
\eqno(1.6)
$$
\par
c) $E(\gamma ,\gamma ')$ is a multivalued form in $\Gamma$ with the following
periodic conditions:  $E(\gamma +a_{k},\gamma '{}) = E(\gamma ,\gamma '{})$, 
$E(\gamma +b_{k},\gamma '{}) = \pm  E(\gamma ,\gamma '{}) 
\exp(-\pi iB_{kk}+ 2\pi i\int ^{\gamma '{}}_{\gamma }{\bf w}_{k})$ where 
$a_{k}$ and $b_{k}$ are the basic cycles.
\par
4) Meromorphic differentials $\Omega_{k}$ and $dp_{k}$.  Let us have a fixed
point $\infty$ in $\Gamma $ with a local parameter $z$. Then we introduce
differentials $\Omega _{k}$ and $dp_{k}$  with the unique pole at $\infty $ 
such that $\Omega _{k}= d(1/z^{k})+O(1)$, $i dp_{k}=d(1/z^{k})+O(1)$ and
$$
\oint _{a_{l}}\Omega _{k}= 0,\ {\rm Im}\, \oint _{a_{l}}dp_{k}= 0, 
\ {\rm Im}\,\oint _{b_{l}}dp_{k}= 0,\ l=1,\ldots g.    
\eqno(1.7)
$$
The multivalued functions $p_{k}$ are called quasimomentums \cite{16}. In the
soliton theory they correspond to the times $t_{k}$. The functions 
${\rm Im}\,p_{k}$ are correctly defined on $\Gamma $. 
For $\Omega _{k}$ we have:
$$
d_{\gamma }d_{z} \ln E(\gamma ,z)= -\sum^{\infty }_{1} \Omega _{k}(\gamma )
z^{k-1} dz .
\eqno(1.8)
$$
\par 
5) In the neighbourhood of $\infty $ we have the following expansions:
$$
\ln\frac{E(z,z')}{z-z'} =
\sum_{m\ge 2} \frac{Q_{m0} (z^{m}+(z')^{m})}{m} +
\sum_{m,n\ge 1} \frac{Q_{mn}z^{m}(z')^{n}}{mn},
\ Q_{mn}=Q_{nm}, 
\eqno(1.9)
$$
$$
\Omega _{k} =d\left(\frac{1}{z^{k}}\right)-\sum_{m\ge 1}Q_{km}z^{m-1}dz ,
\ k\ge 1 ,
\eqno(1.10)
$$
$$\vec{\bf w} = -\sum_{k\ge 1}\vec{U}_{k} z^{k-1}dz ,\ \hbox{where} \ 
(\vec{U}_{k})_{m}= (2\pi i)^{-1}\oint _{b_{m}}\Omega _{k}.
\eqno(1.11)
$$
\par
1.3. {\em Divisors and holomorphic bundles on the Riemann surfaces}.
Let us recall some necessary constructions from algebraic geometry. A
divisor is a formal linear combination of points of the Riemann
surface $\Gamma$: $D=\sum  n_{i}\gamma _{i}$.  If $f$ is a meromorphic 
function on $\Gamma$ with poles $\gamma _{i}$ of order $m_{i}$ and zeroes 
$\gamma ^{*}_{i}$of order $n_{i}$, then the divisor $D(f)=
\sum -m_{i}\gamma _{i}+n_{i}\gamma ^{*}_{i}$
corresponds to it.  Two divisors $D$ and $D'$ are called
equivalent if $D-D'{}$ is a divisor of some meromorphic function.  The
classes of equivalent divisors form the Picard group ${\rm Pic}\,(\Gamma )$.  
Let $D_{1}=\sum  n_{i}\gamma _{i}$, $D_{2}=\sum  \tilde{n}_{i}\gamma _{i}$ 
(some of $n_{i}, \tilde{n}_{i}$ can be equal to 0).  Then $D_{1}\ge D_{2}$ if
$n_{i}\ge m_{i}$ for all $i$.  The sum $\deg(D)=\sum -m_{i}+n_{i}$ 
is called the degree of $D$.
\par
A holomorphic bundle is a bundle with a holomorphic gluing law.
It is very convenient to describe one-dimensional holomorphic bundles
via divisors.  Let ${\Bbb B}$ be a one-dimensional holomorphic bundle, 
$s(\gamma )$ be its global meromorphic section, $D(s)$ be the divisor of $s$, 
and $b$ be the element of $\,{\rm Pic}\,(\Gamma )$ generated by $D(s)$. 
The divisor $D(s)$ depends of course on the section $s(\gamma )$; but 
different sections result in equivalent divisors, so the map 
${\Bbb B} \rightarrow  b \in  \,{\rm Pic}\,(\Gamma )$ is correctly defined.
In algebraic geometry the following statement is well-known:
\par
{\sl Lemma 1.1}. The map from the set of one-dimensional holomorphic 
bundles on $\Gamma $ to
the ${\rm Pic}\,(\Gamma )$ group is a one to one correspondence.
\par
Let ${\Bbb B}$ be a holomorphic bundle on $\Gamma $ with a global meromorphic
section $s$ -- the so-called equipped bundle; $U$ be a domain in 
$\Gamma $; and $D'{}(s)$ be the restriction of $D(s)$ on U.  
Then the holomorphic sections $t$ of ${\Bbb B}$ on $U$ can 
be represented by the meromorphic functions $f$ in $U$
such that $D(f)+D'{}(s)\ge 0$ in the following way: $t=f\cdot $s. 
The meromorphic sections with divisor $D(t)\ge D_{0}$ correspond 
to meromorphic functions such that $D(f)+D'{}(s)\ge D_{0}$. 
Thus we can speak about meromorphic functions with prescribed 
singularities instead of holomorphic sections of bundles.
\par
If the section $s(\gamma )$ has no zeroes and poles in $U$, we have a
trivialization of ${\Bbb B}$ on $U$ and $s(\gamma )$ is called a unit section.
\par
We shall also speak about meromorphic $j$-tensors with a given set
of zeroes and poles. Such bundles are isomorphic to bundles of
0-forms with different divisors. But we shall not use this
isomorphism because we need to vary the basic curve $\Gamma $.  If we vary
the basic curve it is necessary to describe how the bundles are
varied and this variation will depend on the tensor weight j.
\par
The multidimensional bundles are not considered here.
\par
1.4.{\em Deformations of Riemann surfaces and the Riemann problem}.
In this section we consider how the algebra of the vector  fields  on
the circle varies the structures of Riemann surfaces \cite{17}. Let $S$ be a
small circle around  $\infty $  on  $\Gamma $  and  $U(S)$  be its small 
neighbourhood such that $\infty \in U(S)$. Let $\Gamma $ be covered by 
two regions  $\Gamma _{+}$ and  $\Gamma _{-}$ such that  
$U(S)=\Gamma _{+}\cap \Gamma _{-}$ and  $\infty \in \Gamma _{-}$. $\Gamma$ 
may by treated as a result of gluing  $\Gamma_{+}$  and $\Gamma_{-}$.
We may vary the Riemann surface  $\Gamma$ by changing the gluing law. 
Let us describe this change.  Let  $v=\tilde v(z )d/dz$  be  a
holomorphic vector field in the region  $U(S)$, and  
$\exp(\beta v)\gamma _{-}$  be the shift of the point  $\gamma_{-}$  
along  $v$  after the lapse of time $\beta $. Let the original gluing law 
be $\gamma _{+}\rightarrow \gamma _{-}$, $\gamma _{-}\in \Gamma _{-}$, 
$\gamma _{+}\in \Gamma _{+}$. Now we obtain a new Riemann surface 
$\Gamma^{'}$  by gluing the point $\gamma_{+}$ to the point
$\exp(\beta v)\gamma _{-}$.  Both Riemann surfaces $\Gamma$ and 
$\Gamma^{'}$ are constructed of the same regions  $\Gamma _{+}$ and 
$\Gamma_{-}$. Then the unit maps $\Gamma _{+}\rightarrow \Gamma _{+}$, 
$\Gamma_{-}\rightarrow \Gamma _{-}$ define a natural mapping 
$\Gamma \rightarrow \Gamma ^{'{}}$ with a jump on $S$. We call this mapping
$E$. When calculating the commutators of vector field actions we assume
all vector fields  to  be  independent  of  $\beta$ functions of local
parameter $z=1/\lambda,z$ to be defined on $\Gamma _{-}$, and when we 
vary Riemann surfaces we map $z$ by $E$.
\par
Let $D_{0}$ be a divisor on $\Gamma$ and ${\Bbb B}_{j}$ be the bundle of 
$j$-tensors $f_{j}$ such that $D(f_{j})\ge -D_{0}$. We assume that the 
corresponding bundle ${\Bbb B}^{'}_{j}$ on the new surface $\Gamma^{'}$ 
is the bundle of $j$-tensors $f^{'{}}_{j}$ such that 
$D(f^{'{}}_{j})\ge -E(D_{0}).$
\par
In the case of infinitesimal action $(\beta <<1)$, a holomorphic
$j$-tensor field $\Delta '{}$ on the new surface $\Gamma ^{'}$ can 
be treated as a field on the old surface $\Gamma$ with a jump on 
$S$ satisfying the following equation
$$
\Delta{'}_+ - \Delta{'}_- = \beta  L_{v}\Delta 
\eqno(1.12)
$$
where $\Delta{'}_+ $ and $\Delta {'}_-$ are the boundary values of 
$\Delta'{}$ on $S$, $L_{v}$ is the Lie derivative, and $\Delta$ is the 
original field on $\Gamma $. Thus $\Delta '{}$ is a solution
of the Riemann problem -- a well-known object of soliton theory. We
assume in our paper that the index of (1.12) is zero. The Riemann
problem (1.12) can be considered as a special case of the 
$\bar{\partial}$-problem (1.1) with a $\delta$-type function $g$. 
Then it's solution is given by:
$$
\Delta'{}(\gamma) =\Delta(\gamma) + \oint _{S}\omega (\gamma ,\gamma '{}) 
\beta L_{v}\Delta (\gamma '{}) ,
\eqno(1.13)
$$
where $\omega (\gamma ,\gamma '{})$ is the same Cauchy kernel as in the 
$\bar{\partial}$-problem (the kernel $\omega (\gamma ,\gamma '{})$ depends, 
of course, on the bundle ${\Bbb B}_{j})$. For example a calculation for 
a holomorphic 1-form using (1.13) gives rise to the
well-known formula for the variation of Riemann matrix  $B_{mn}$ \cite{17}:
$$
\frac{\partial B_{mn}}{\partial \beta} =
\oint _{S}\tilde v(z)\tilde{\bf w}_{m}(z)\tilde{\bf w}_{n}(z)dz ,
\eqno(1.14)
$$
where $\tilde{\bf w}_{m}(z)dz$, $m=1,\ldots ,g$ is the basis of 
holomorphic 1-forms.
\par
\medskip
\noindent {\bf Chapter 2. Elements of the Kadomtsev-Petviashvili equation
theory.}
\par
\medskip
In this chapter the necessary definitions are introduced. The
most important of them are the Baker-Akhiezer functions, the
Baker-Akhiezer $j$-tensors and the Segal-Wilson $\tau $-function. The main
tool is the Cauchy-Baker-Akhiezer kernel which inverts the $\bar{\partial}$
operator. The first section illustrates the further considerations
and may be omitted.
\par
\medskip
2.1 {\em Riemann surfaces in the integrable equations theory}.  Now we
shall show how the Riemann surfaces and the Baker-Akhiezer function
appear in the theory of nonlinear equations \cite{18}. Then we shall show
that the Virasoro action corresponds to higher symmetries of these
equations. Consider the simplest example -- the Korteveg-de Vries
equation (KdV):
$$
u_{t}-6uu_{x}+u_{xxx}=0.
\eqno(2.1)
$$
It is a Hamiltonian equation
$$
u_{t}=\frac{d}{dx}\frac{\delta H}{\delta u}, \  
H=\int \left(\frac12 u^{2}_{x}+ u^{3} \right) dx
\eqno(2.2)
$$
with an infinite set $H_{1},H_{2},\ldots $ of first integrals in involution, 
$H_{n}=\int h_{n}(u,u_{x},\ldots ) dx$.
{\sloppy

}
The scheme of solving this equation is based upon the following
representation for KdV (Lax representation). Let 
$L =-\partial ^{2}/\partial x^{2}+ u(x,t)$, $A = \partial /\partial t + 
4\partial ^{3}/\partial x^{3}- 6u \partial /\partial x- 3u_{x}$. 
Then operators $L$ and $A$ commute
$$
LA=AL
\eqno(2.3)
$$
if and only if $u(x,t)$ is a solution of (2.1). One can see from (2.3)
that the spectrum of $L$ is independent of $t$. (If we speak about the
spectral properties of $L$, we consider it as an ordinary differential
operator depending on a parameter $t$). If $u(x,t)$ is periodic in $x$:
$u(x+T,t)=u(x,t)$ then the spectrum of $L$ consists of a set of intervals
$[E_{0},E_{1}]$, $[E_{2},E_{3}]$, $[E_{4},E_{5}],\ldots$ 
$E_{0}< E_{1}< E_{2}<\ldots $
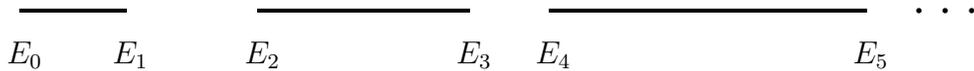
\par\noindent
\begin{picture}(380,50)
\thicklines
\put(15,30){\line(1,0){40}}
\put(10,10){$E_0$}
\put(50,10){$E_1$}
\put(105,30){\line(1,0){80}}
\put(100,10){$E_2$}
\put(180,10){$E_3$}
\put(215,30){\line(1,0){120}}
\put(210,10){$E_4$}
\put(330,10){$E_5$}
\put(355,30){\circle*{2}}
\put(365,30){\circle*{2}}
\put(375,30){\circle*{2}}
\end{picture}
\par
The open intervals $(-\infty ,E_{0})$, $(E_{1},E_{2}),\ldots $ are called 
gaps, $E_{2n}-E_{2n-1}\rightarrow 0$ as $n \rightarrow \infty$.  
The Bloch eigenfunction of the operator $L$ (it coincides with the 
Baker-Akhiezer function in this case) is the solution of
$$
L \psi (x,E,t)= E \psi (x,E,t)
\eqno(2.4)
$$
such that $\psi (x+T,E)=\exp(i T p(E)) \psi (x,E,t)$ with the normalization
$\psi(0,E,t)=1$. The function $\psi (x,E,t)$ is meromorphic in $E$ on a 
Riemann surface $\Gamma$ which is two-sheeted over the $E$-plane; the 
branch points are $E_{0}$, $E_{1},\ldots $ . The function 
$\psi (x,E)$ has exactly one pole $\gamma _{n}(t)$ and one zero 
$\gamma ^{+}_{n}(x,t)$ over each gap except $(-\infty ,E_{0})$; 
$\gamma ^{+}_{n}=\gamma _{n}$ as $x=0$.
{\sloppy

}
If we know $E_{n}$ and $\gamma _{n}(t)$, then the function  $u(x,t)$ can be
reconstructed in all x. The shift along the flow corresponding to the
Hamiltonian $H$ as well as $H_{m}$ results in change in the positions of the
poles $\gamma_{n}$, the points $E_{n}$ being invariant. Finite-dimensional 
invariant subspaces correspond to the so-called finite-gap potentials, i.e., 
to the potentials such that the spectrum has the form
$[E_{0},E_{1}]$, $\ldots$, $[E_{2N},\infty ]$. These potentials are the 
stationary points of equations with Hamiltonians of the form  
${\bf H} =\sum_{m=1}^N c_{m}H_{m}$, $\frac{d}{dx} 
\frac{\delta {\bf H}}{\delta u}=0$.
The restriction of the KdV equation on this subspace gives rise
to finite-dimensional systems that are integrable in the Liouville
sense.  Roughly speaking, the action variables correspond to the
sizes of gaps and the angle variables correspond to the positions of
$\gamma $ on the Riemann surface $\Gamma$.
\par
Except for the symmetries corresponding to the Hamiltonians $H_{m}$, 
other flows commuting with KdV equation do exist \cite{19}.  They can be
written in the following form:
$$
\frac{\partial u}{\partial\beta_m}=
\frac{d}{dx} \Lambda ^{m+1}\cdot (6tu + x),
\eqno(2.5)
$$
where
$$
\Lambda = -\left(\frac{d}{dx}\right)^2 + 4\left(\frac{d}{dx}\right)^{-1} 
u \left(\frac{d}{dx}\right)+ 2\left(\frac{d}{dx}\right)^{-1} u_{x}
\eqno{(2.6)}
$$
is called a recursion operator \cite{20}.
\par
Ordinary symmetries of the KdV equation which correspond to the
Hamiltonians $H_{m}$, also can be written via $\Lambda$ :
$$
\frac{\partial u}{\partial t_m}=
\frac{d}{dx} \Lambda ^{m}\cdot \frac{1}{2}.                      
\eqno(2.7)
$$
\par
In our paper we study how equations (2.5) act on the finite-gap
KdV solutions. The ends of gaps are not invariant under these flows,
but:
$$
\frac{\partial E_{k}}{\partial \beta _{m}}= E^{m+1}_{k}.
\eqno(2.8)
$$
\par
In particular flows $\partial E_{k}/\partial \beta _{-1}=1$, 
$\partial E_{k}/\partial \beta _{0}=E_{k}$, 
$\partial E_{k}/\partial \beta _{1}=E^{2}_{k}$ which represent 
infinitesimal fractional transformations of the $E$-plane,
correspond to:
$$
\partial u/\partial \beta _{-1}=6tu_{x}+ 1\qquad 
\hbox{(Galilean transformation)},
$$
$$
\partial u/\partial \beta _{0}=6tu_{t}+ 2xu_{x}+ 4u\qquad 
\hbox{(Scaling transformation)},
$$
$$
\partial u/\partial \beta _{1}=6t(u_{xxxx}-10uu_{xx}-5u^{2}_{x}+
10u^{3})_{x}+2xu_{t}+16u^{2}+4u_{x}\left(\frac{d}{dx}\right)^{-1}u.
$$
\par
The flows (2.5) commute as the corresponding vector fields
$E^{m+1}\partial /\partial E$ (see (2.8)).
\par
The shift (2.8) of the branch points $E_{k}$ of the surface 
$\Gamma$ generates the variation of the complex structure of $\Gamma$. 
As one can see, vector fields which do not move $E_{k}$ appear to be
generalized-graded \cite{7}.
\par
In the KdV theory only the  hyperelliptic  Riemann  surfaces
emerge.  Arbitrary Riemann surfaces appear in the theory  of  the  KP
equation:
$$
(4u_{t}- u_{xxx}- 6uu_{x})_{x}= 3u_{yy}.
\eqno(2.9)
$$
\par

The fact that the scaling transformation can be obtained from the
Galilean one by applying the recursion operator was first pointed out
in \cite{19a}. In \cite{19a} it was also shown that applying the recursion
operator to the scaling symmetry we get a non-local KdV symmetry.
But in \cite{19a} only the local symmetries were studied, thus the last
observation had no consequences in the context of \cite{19a}.

2.2  {\em The Krichever construction. The Baker-Akhiezer function.}
(see review \cite{16}). Let $\Gamma$ be a Riemann surface of genus $g<\infty$ 
with a given point $\infty$, a local parameter $z = 1/\lambda $ 
in the neighbourhood of $\infty$, and a divisor $\gamma _{1},\ldots, 
\gamma _{g}$. The Baker-Akhiezer function $\psi (\gamma ,\vec{t})$ is a
function uniquely determined by the following properties. 
\par
1) It
depends on a spectral parameter $\gamma \in \Gamma$ and an infinite 
set of times $\vec{t}=(t_{1},\ldots  ), t_{1}=x, t_{2}=y, t_{3}=t$. 
\par
2) It is meromorphic by $\gamma $ everywhere
but $\infty $, and has simple poles in $\gamma _{1},\ldots  ,\gamma _{g}$. 
\par
3) It has an essential singularity: 
$$
\psi (\gamma ,\vec{t})= 
(1-\chi (\vec{t})/\lambda -o(1/\lambda ))\exp(\sum  \lambda ^{m}t_{m}),\ 
\lambda =\lambda (\gamma )\ \hbox{as}\ \gamma \rightarrow \infty. 
$$  
\par
We call $\chi$ potential; it obeys the KP equation: 
$$
4\chi _{xt}=\chi _{xxxx}+6\chi _{x}\chi_{xx}+3\chi_{yy}.
$$
(In (2.9) $u = \chi _{x}$.) The $t_{m}$- dependence of $\chi$  ($m >3$) 
is described by the higher KP equations (the so-called KP-hierarchy). 
We need also a conjugated Baker-Akhiezer differential $\psi ^{*}(\gamma ,t)$, 
which is holomorphic by $\gamma$  1-form on $\Gamma\backslash\infty$. 
$\psi ^{*}(\gamma ,t)$ has simple zeroes at $\gamma _{1},\ldots  ,\gamma _{g}$
and an essential singularity  
$$
\psi ^{*}(\lambda ,\vec{t})=(1+O(1/\lambda ))
\exp(-\sum  \lambda ^{m}t_{m}) d\lambda\ \hbox{as}\ 
\lambda \rightarrow \infty.
$$
(for explicit formulas for $\psi (\gamma ,\vec{t})$ and 
$\psi ^{*}(\gamma ,\vec{t})$ see 2.3). We also need:
\par
{\sl Lemma 2.1} (see \cite{16}).  Let  ${\rm Im}\, p_{1}(\lambda )= 
\,{\rm Im}\, p_{1}(\mu )$ (for quasimomentum $p_{1}$, see 1.2). 
Then the functions  $\psi $ and $\psi^{*}$ are orthogonal functions of $x$: 
$$
\int ^{+\infty }_{-\infty }\psi (\lambda ,x,y,t,\ldots  ) 
\psi ^{*}(\mu ,x,y,t,\ldots  ) dx =0\ \hbox{as}\  \lambda \neq \mu.
$$
\par
2.3. {\em Baker-Akhiezer j-forms. The bundles ${\Bbb B}_{j}(t_{0},\vec t,D)$.}
Along with the Baker-Akhiezer functions we shall use Baker-Akhiezer
$j$-differentials $\psi _{j}(\gamma ,t_{0},\vec{t})$, introduced in
\cite{7},\cite{9}  
(Baker-Akhiezer 1-differentials were introduced in \cite{21}).
\par
Let $D = \sum  n_{m}\gamma _{m}$ be some divisor of degree $(2j-1)(g-1)$ 
(see 1.3). We shall consider the following equipped holomorphic bundle
${\Bbb B}_{j}(t_{0},\vec{t},D)$: local holomorphic sections of 
${\Bbb B}_{j}(t_{0},\vec{t},D)$ are meromorphic
on $\Gamma \backslash \infty$  $j$-forms $f(\gamma)$ such that 
$D(f) \ge  D+(t_{0}-1)P_{0}$ and 
$f(z)(z)^{t_{0}-1}(dz)^{-j}\exp -\sum  z^{-m}t_{m}$  is regular at 
$\infty$ $(z=1/\lambda )$. The index of
$\bar{\partial }_{j}$ on ${\Bbb B}_{j}(t_{0},\vec{t},D)$ equals 0. 
Thus we eliminate the
$(2j-1)(g-1)$-dimensional kernel of $\bar{\partial }_{j}$ on the trivial 
bundle. For $\vec{t}=0$, $t_{0}=1$ we denote 
${\Bbb B}_{j}(t_{0},\vec{t},D)={\Bbb B}_{j}$. The conjugate bundle to 
${\Bbb B}_{j}(t_{0},\vec{t},D)$ is
${\Bbb B}^{*}_{j}(t_{0},\vec{t},D)={\Bbb B}_{1-j}(2-t_{0},-\vec{t},-D)$ 
(i.e, the bundle  ${\Bbb B}^{*}_{j}(t_{0},\vec{t},D)$ consists of
the $1-j$-differentials).
\par
The Baker-Akhiezer $j$-differential $\psi _{j}(\gamma ,t_{0},\vec{t})$ 
and the conjugate $1-j$-differential 
$\psi ^{*}_{j}(\gamma ,t_{0},\vec{t})$ can be defined as 
meromorphic sections of
${\Bbb B}_{j}(t_{0},\vec{t},D)$ and  ${\Bbb B}^{*}_{j}(t_{0},\vec{t},D)$, 
respectively, with no singularities except simple poles in $\infty $.   
For $\psi _{j}(\gamma ,t_{0},\vec{t})$ we have the explicit formula:
$$
\psi _{j}(\gamma ,t_{0},\vec{t}) = \eta (\gamma ) \psi _{1/2}
(\gamma ,t_{0},\vec{t}), \ \hbox{where}
\eqno(2.10)
$$
$$
\eta (\gamma ) = \left[\frac{\theta(\vec A(\gamma)-\vec K)}
{\theta^{(g)}(-\vec K) E(\gamma,\infty)\sqrt{dz(\infty)}}                  
\right]^{2j-1} \prod_k \left[\frac{E(\gamma,\gamma_k)}
{E(\gamma,\infty)E(\infty,\gamma_k) dz(\infty)}\right]^{n_k},
$$
$$
\psi _{1/2}(\gamma ,t_{0},\vec{t})= \left[\frac{E(\gamma,P_0)}
{E(\gamma,\infty)E(\infty,P_0) dz(\infty)}\right]^{t_0-1}
\frac{\theta(\vec A(\gamma)+\vec\xi)}
{\theta(\vec\xi) E(\gamma,\infty)\sqrt{dz(\infty)}}
\exp\left(\int\limits^{\gamma}\sum _{k\ge 1}\Omega_{k}t_{k} \right),
$$
$$
\theta ^{(g)}(-\vec{K})=\frac{1}{g!}\frac{d^g}{dz^g}
\left.\theta (\vec{A}(z)-\vec{K})\right|_{z=0},
$$
$$ 
\vec{\xi }=(2j-1)\vec{K}+\sum  n_{k}\vec{A}(\gamma _{k})+(t_{0}-1)\vec{U}_{0}+
\sum _{k\ge 1}\vec{t}_{k}\vec{U}_{k}+\vec{\zeta},
$$
where $\vec{K}$ is the vector of Riemann constants; 
$\vec{U}_{0}=\vec{A}(P_{0})$: for Abel
transform  $\vec{A}$  see  1.2; $\vec{\zeta}=0$;  for 
$\theta$-functions see \cite{14}. The constants in the integrals in (2.10) are 
chosen so that $\int^{\gamma}\Omega _{k}=1/z^{k}+o(1)$.  When  
$\gamma \rightarrow \infty$ , $\eta (\gamma )\sim(dz)^{j-1/2}$.
\par
If  $\vec{\zeta }\neq 0$ (2.10) results in Baker-Akhiezer functions with 
nonzero characteristics: $\psi _{j}(\gamma +a_{k},t_{0},\vec{t})=\psi _{j}
(\gamma ,t_{0},\vec{t})$, $\psi _{j}(\gamma +b_{k},t_{0},\vec{t}) =
\psi _{j}(\gamma ,t_{0},\vec{t}) \cdot \exp(2\pi i\zeta _{k})$ 
where $a_{k}$ and $b_{k}$ are basis cycles.
\par
Not only integer $j$ but $j\in {\Bbb Z}/2$ can be considered (see \cite{9}).
\par
{\it Remark.} For $j=1/2$ we can take $D=0$ and parameterize the
Baker-Akhiezer functions by characteristics $\vec{\zeta }.
$\par
The functions $\psi _{j}(\gamma ,t_{0},\vec{t})$ and 
$\psi ^{*}_{j}(\gamma ,t_{0},\vec{t})$ have the following
asymptotics, when $\gamma \rightarrow \infty$ :
$$
\psi_{j} (\gamma,t_0,\vec{t})= z^{-t_0}(dz)^j \exp \left(\sum_{k\ge1} z^{-k}
t_k \right ) (1+O(z)), 
$$
$$
\psi^{*}_{j}(\gamma,t_0,\vec{t})= z^{t_0-2} (dz)^{1-j} \exp \left(-\sum_{k\ge1}
z^{-k} t_k \right ) (1+O(z)), 
\eqno(2.11)
$$
For  $\gamma  \rightarrow  P_{0}$  we  have:
$$
\psi_{j} (\gamma,t_0,\vec{t}) \sim \varphi (t_{0},\vec{t}) z^{t_0-1} (dz)^j, 
\ \ \ 
\psi^{*}_{j}(\gamma,t_0,\vec{t}) \sim \varphi^* (t_{0},\vec{t}) z^{1-t_0} 
(dz)^{1-j}
\eqno(2.12)
$$
where $z_{-}$ is some local parameter in $P_{0}$.
\par
The variations of the Baker-Akhiezer forms are solutions of the
corresponding Riemann problem in ${\Bbb B}_{j}(t_{0},\vec{t},D)$ (see 3.2).
\par
For all $j$, the bilinear identity \cite{3} 
$\oint _{S}\psi _{j}(\gamma ,t_{0},\vec{t})\psi ^{*}_{j}
(\gamma ,t'_{0},\vec{t}'{})\equiv 0$
for $t_{0}\ge t'_0$ holds. For $t_{0}=t'_0$ it results in the 
ordinary KP-hierarchy.
The corresponding solutions do not depend on $j$. However, for the
equations corresponding to the changing complex structure, the tensor
weight is important.
\par
\noindent From (2.11) we have the orthogonality condition:
$$
\oint _{S}\psi _{j}(\gamma ,t_{0},\vec{t})
\psi ^{*}_{j}(\gamma ,t'_0,\vec{t}) = -\delta (t_{0}+1,t'_0 ),
$$
which  was derived for $\vec{t}=0$ in \cite{7}. Let us note that 
$\psi _{j}(t_{0})$ and $\psi ^{*}_{j}(t'_0)$
form full mutually orthogonal bases (see \cite{7}) as functions of $\gamma$.
\par
{\it Remark.} If all points of $\gamma _{k}$ coincide with $P_{0}$ or 
$\infty $ then for $\vec{t}=0$ we obtain the Krichever-Novikov 
bases for $j$-forms \cite{7}.
\par
2.4. {\em ``Cauchy-Baker-Akhiezer'' kernel}. We shall solve the
Riemann problem (1.12) and the $\bar{\partial }$-problem for  the  
bundle  ${\Bbb B}_{j}(t_{0},\vec{t},D)$
with the help of the Cauchy-Baker-Akhiezer kernel:
$$
\omega _{j}(\lambda,\mu,t_{0},x,y,t,\ldots)=\int\limits^{x}_{\mp\infty} 
\psi _{j}(\lambda,t_{0},x',y,t,\ldots) 
\psi^{*}_{j}(\mu,t_{0},x',y,t,\ldots) \frac{dx'}{2\pi i}.
\eqno(2.13)
$$
We choose the sign so that the integral converges. For 
${\rm Im}\, p_{1}(\lambda )= {\rm Im}\,p_{1}(\mu)$ 
the definition (2.13) is correct because of the Lemma 2.1.
\par
One can check that $\psi _{j}(\gamma ,t_{0}+1,\vec{t}) = 
(\partial/\partial t_{1}-v_{j}(t_{0},\vec{t}))\psi _{j}(\gamma ,t_{0},
\vec{t})$ and $\psi ^{*}_{j}(\gamma ,t_{0},\vec{t}) = 
(\partial /\partial t_{1}+v_{j}(t_{0},\vec{t})) 
\psi ^{*}_{j}(\gamma ,t_{0}+1,\vec{t}), v_{j}(t_{0},\vec{t})=
\partial /\partial t_{1}\ln(\varphi (t_{0},\vec{t}))$
(see 2.12). Therefore we have a different representation for the
Cauchy-Baker-Akhiezer kernel:
$$
\omega _{j}(\gamma ,\gamma '{},t_{0},\vec{t}) =\frac{1}{2\pi i} 
\left(\sum^{t_0}_{-\infty }\ \hbox{or}\ \sum^{+\infty }_{t_0+1}\right) 
\psi _{j}(\gamma ,t'_0-1,\vec{t}) \psi ^{*}_{j}(\gamma '{},t'_0,\vec{t}), 
\eqno(2.14)
$$
which was obtained for $\vec{t}=0$ earlier in \cite{8}.
\par
{\sl Lemma 2.2}. For $\lambda,\mu\neq\gamma_{k},P_{0},\infty$
we have:
$$
\frac{\partial}{\partial \bar{\lambda}} 
\omega _{j}(\lambda ,\mu ,t_{0},\vec{t}) = 
\delta (\lambda -\mu ) d\mu d\bar{\mu },
\eqno(2.15)
$$
where $\delta (\lambda -\mu )$ is the two-dimensional $\delta $-function. 
For $t_{0}=0$, $\vec{t}=0$ we obtain a new representation for the known 
Cauchy kernel on the Riemann surface (see 1.1).
\par
If the operator $\partial /\partial \bar{\lambda }$ acts on the bundle 
${\Bbb B}_{j}(t_{0},\vec{t},D)$ then (2.15) is valid for all $\lambda$, $\mu$.
\par
We also use the "vacuum" Cauchy kernel corresponding to 
$\Gamma ={\Bbb C}{\Bbb P}^{1}$,$P_{0}=0$:
$$
\omega ^{0}_{j}(z,z'{},t_{0},\vec{t}) = (z/z'{})^{1-t_{0}}(z'{}-z)^{-1}
(dz)^{j}(dz')^{1-j}/2\pi i. 
\eqno(2.16)
$$
\par
2.5. {\em Grassmannians and flag spaces corresponding to $j$-forms}.
Grassmannians in soliton theory and the $\tau $-function were introduced in
papers by the Kyoto group (see $\cite{28},\cite{3}$ and references
therein). Here
we use the approach by G.Segal and G.Wilson (see \cite{22})  but  with  an
extra discrete parameter $t_{0}$ dependence. The tensor properties are not
discussed in these works. The Grassmannians of $j$-differentials and
Virasoro action on them were treated in \cite{2}. Now we use the flag
spaces instead of Grassmannians. Let $S$ be a small contour in the
neighborhood of $\infty $ such that there are no points of $D$ inside 
$S$, and $z=1/\lambda $ be a local parameter at $\infty $. 
Let $H={\cal L}^{2}(S)$ be the space of square
integrable $j$-forms on S. For each $t_{0}$ let us have a decomposition
$H=H_{+}(t_{0}) \oplus  H_{-}(t_{0})$, where $H_{+}(t_{0})$ and 
$H_{-}(t_{0})$ are  subspaces, generated by the basis elements 
$z^{i}(dz)^{j}$  with  $i < -t_{0}$  and  $i \ge  -t_{0}$,
respectively.  The flag space $Fl_{j}$ is the set of  stratified  families
of subspaces $W=\{W(t_{0})\}$  such  that  $W(t_{0})\subset W(t_{0}+1)$  
for  all  $t_{0}$,  the orthogonal projections 
$P_{+}(t_{0}): W(t_{0}) \rightarrow  H_{+}(t_{0})$ are Fredholm  operators
of index 0 and the projections $P_{-}(t_{0}): W(t_{0}) \rightarrow  
H_{-}(t_{0})$  are  compact. Each $W(t_{0})$ is an element of the 
corresponding  Grassmannian.  Let
$w=\{e_{k}\}$ be a basis in $H$ such that $w(t_{0})=\{ e_{k}\mid k <-t_{0}\}$ 
is a basis  in $W(t_{0})$.  Let $w_{+}(t_{0})=P_{+}(t_{0})w(t_{0}), 
w_{-}(t_{0})=P_{-}(t_{0})w(t_{0})$,  and   $A(t_{0})=
P_{-}(t_{0})P^{-1}_{+}(t_{0})$. It is convenient to write $w$ as a block  
matrix  whose columns correspond to the Laurent expansions of $e_{k}$:
$$
w=\left[ \begin{array}{c}w_+ \\ w_- \end{array}\right],
\ A(t_{0})= w_{-}(t_{0})(w_{+}(t_{0}))^{-1}.
\eqno (2.17)
$$
The elements $W(t_{0})$ which correspond to the given Baker-Akhiezer
function $\psi _{j}(\gamma ,t_{0},\vec{t})$ consist of all sections 
$f(\gamma )$ of  ${\Bbb B}_{j}(t_{0},\vec{t},D)$,
which are holomorphic in $\Gamma _{+}$. So  $\psi _{j}(\gamma ,t_{0},\vec{t})$ 
with different $\vec{t}$ generate $W(t_{0}).
$\par
2.6. {\em Transformations of the flag spaces and $\tau$-function}. Let 
$W \in \hbox{Fl}_{j}$, $W=\{W(t_{0})\}$, and $g$ be some linear operator on 
$H$ from $GL(\infty )$. Then the element 
$g^{-1}W=\{g^{-1}W(t_{0})\} \in \hbox{Fl}_{j}$ is the collection of spaces 
$g^{-1}W(t_{0})$ generated by the $j$-forms $gf(z)$, where $f(z)\in W(t_{0})$. 
Continuous functions act on $H$ by ordinary multiplication, and vector 
fields on $S$ act on $H$ by Lie differentiation \cite{2}. We also write
$W(\vec{t})=\exp(-\sum z^{-k}t_{k})W$.
\par
{\it Remark.} There exists some function $\alpha (\gamma )$ such that 
$\alpha W(t_{0}) =W(t_{0}+1)$ for all $t_{0}$, but we do not use this fact now.
\par
Let $t_{0}$ be a fixed number, $g^{-1}=\left[\begin{array}{cc} 
a & b \\ c & d \end{array}\right]$ be a transformation of $H$
(the block form corresponds to the splitting 
$H=H_{+}(t_{0})\oplus H_{-}(t_{0})$, $a=(g^{-1})_{++}$, $b=(g^{-1})_{+-}$, 
$c=(g^{-1})_{-+}$, $d=(g^{-1})_{--})$. Then the $\tau$-function
$\tau _{W}(t_{0},g)$ is determined by the formula:
$$
\tau _{W}(t_{0},g)=(\det(1+a^{-1}bA))\tau _{W}(t_{0},1), \ 
\tau _{j}(t_{0},\vec{t})=\tau _{W}(t_{0},\exp(\sum z^{-k}t_{k})). 
\eqno(2.18)
$$
\par
There is no canonical choice of $\tau _{W}(1)$ (we omit $t_{0}$-dependence in
our notations for the sake of brevity), so $\tau _{W}$ is defined up to a
constant factor. In \cite{22} $\tau _{W}(1)=1$. But if we deform the Riemann
surface by a vector field, it is more natural to assume that the
variation of $\tau _{W}(g)$ is given by (2.18). The composition formula for a
product of transformations is:
$$
\tau _{W}(g\cdot g_{1})= \tau _{W}(g)\cdot  \tau _{(g^{-1}W)}(g_{1})/
\tau _{(g^{-1}W)}(1)\cdot \rho (g,g_{1}),
\eqno(2.19)
$$
$$
\rho (g,g_{1})=\det \{g_{++}(g_{1})_{++}(gg_{1})^{-1}_{++}\}.
$$
\par
We will consider the following transformations of the
Grassmannian: the action of the vertex operator in section 2.8 and
the action of the vector fields in section 3.3. To calculate the
corresponding variations of $\tau $-functions we need the explicit form of
$A(t_{0})=P_{-}(t_{0})P^{-1}_{+}(t_{0})$.
\par
{\it Remark.} {\em Grassmannians as an universal moduli space}.
Every Riemann surface $\Gamma $ of finite genus with a bundle and a
local parameter generates a point $W$ of the Grassmannian $\hbox{Gr}_{j}$ 
(we do not discuss $t_{0}$-dependence now and assume $t_{0}=0)$. 
So the Grassmannian can be considered as a universal space including 
the moduli spaces for all genera. But the points of the Grassmannian 
corresponding to the Riemann surfaces have the following property 
(see \cite{12} and references therein). Let $W^{\perp }\in \hbox{Gr}_{j}$ 
be the set of $1-j$-forms $h(z)(dz)^{1-j}$ such that $\oint f(z)h(z)dz=0$ 
for all $f(z)(dz)^{j}\in W (W^{\perp }$ corresponds to
$\psi ^{*}(\gamma ,\vec{t}))$. Then there exists a $1-2j$-form 
$g(z)=\tilde g(z)(dz)^{1-2j}$, $z\in S$, such that $W^{\perp }=g(z)W$ 
(see \cite{2}). For an arbitrary $W\in \hbox{Gr}_{j}$ this property is not
valid. The set of points $W\in \hbox{Gr}_{j}$ such that $W^{\perp }=g(z)W$ 
for some $g(z)$ is called a universal moduli space.
\par
2.7. {\em Explicit version of Segal-Wilson construction}. The main 
operator $A_{j}(t_{0})$ (see 2.17) can be written explicitly via 
Cauchy  kernels (2.13), (2.16): $A_{j}(t_{0})f(\lambda ,\vec{0}) = 
\oint  (\omega _{j}(\lambda ,\mu ,t_{0},\vec{0})-
\omega ^{0}_{j}(\lambda ,\mu ,t_{0},\vec{0})) f(\mu ,\vec{0})$, where 
$f(\lambda )$ is a $j$-differential. For transformed  Grassmannians 
$W(\vec{t})$ we have $A_{j}(t_{0},\vec{t})=$ 
$\exp( \sum (\lambda ^{n}-\mu ^{n})t_{n})A_{j}(t_{0})$. 
\par
2.8. {\em Vertex operators. Cauchy kernel via $\tau $-function}. Consider 
the vertex operator $X_{j}(z)= \exp(\sum z^{-m}t_{m}-t_{0}\ln z)
\cdot \exp(-\sum z^{n}n^{-1}\partial /\partial t_{n}-\partial /
\partial t_{0})(dz)^{j}$ and its adjoint  
$X^{*}_{j}(z)= \exp(t_{0}\ln z-\sum z^{-m}t_{m})\cdot 
\exp(\sum z^{n}n^{-1}\partial /\partial t_{n}+
\partial /\partial t_{0})(dz)^{1-j}$
determined near $\infty $. They satisfy the following commutation
relations:
$$  
X_{j}(z)X^{*}_{j}(u) + X^{*}_{j}(u)X_{j}(z) = -\delta (z-u)du,
$$ 
$$
X_{j}(z)X_{j}(u) + X_{j}(u)X_{j}(z) = 0,
$$
$$ 
X^{*}_{j}(z)X^{*}_{j}(u) + X^{*}_{j}(u)X^{*}_{j}(z)= 0,
$$ 
so $X_{j}(z)$ and $X^{*}_{j}(z)$ represent fermionic operators 
on the sphere. Then we have
$$
X_{j}(z)X^{*}_{j}(u)\cdot \tau _{j}(t_{0},\vec{t}) = 
-2\pi i\omega _{j}(z,u,t_{0},\vec{t})\tau _{j}(t_{0},\vec{t}) .
\eqno(2.20)
$$
The proof follows from the relation  
$$
X_{j}(\lambda )X^{*}_{j}(\mu )\cdot \tau _{W}(g) =
\omega ^{0}_{j}(\lambda ,\mu ,t_{0},\vec{t})\cdot \tau _{W}
(g(1-k/\lambda )/(1-k/\mu )),
$$
where  $g = \exp \sum k^{m}t_{m}$, 
from the composition formula (2.19) and the explicit representation 2.7.
\par
2.9. {\em Algebrogeometrical $\tau$-function for $j$-forms}. The 
algebro-geometrical $\tau$-function without $t_{0}$-dependence was 
discussed in \cite{5},\cite{23},\cite{24},\cite{1},\cite{4}. In 2.6 the 
$\tau$-function was determined up to an arbitrary factor $c(t_{0})$. 
To eliminate this freedom we use the following condition:
$$
\tau _{j}(t_{0}+1,\vec{t}) / \tau _{j}(t_{0},\vec{t}) = 
\varphi (t_{0},\vec{t}),
\eqno(2.21)
$$
$\varphi$ is defined by (2.12).  The equivalent condition for vacuum
expectation \hbox{$<-p\mid p>$} ($p$ being correspondent to $t_{0}$ in 
our paper) was used by I.M.Krichever and S.P.Novikov ((2.34) in \cite{9}). 
Then we have:
$$
\begin{array}{c}
\tau _{j}(t_{0}+1,\vec{t})= \\ \\
= \exp\left\{\frac12\left(\sum_{1}^{\infty} 
Q_{ik}t_{i}t_{k}+ g_{2}t^{2}_{0}+ 2t_{0}\sum_{1}^{\infty}
(Q_{0k}-q_{0k})t_{k}\right) +
\sum_{1}^{\infty} h_{k}t_{k}+g_{1}t_{0}+g_{0} \right\}\cdot \\ \\
\cdot 
\theta \left\{(2j-1)\vec{K} +\sum  n_{k}\vec{A}(\gamma _{k})+
t_{0}\vec{U}_{0}+\sum_{k\ge 1} t_{k}\vec{U}_{k}+\vec{\zeta }
\mid B_{mn}\right\}. 
\end{array}
\eqno(2.22)
$$
Notations are the same as in 1.2, (2.10). The terms $Q_{0k}-q_{0k}$ and $h_{k}$
are defined from decompositions 
$\ln \{(zE(z,P_{0})/ (E(z,\infty )E(\infty ,P_{0})dz(\infty ))\}=
-\sum (Q_{0k}-q_{0k})z^{k}/k, \ln \eta  =-\sum h_{k}z^{k}/k$. 
The terms $g_{2}, g_{1}$ are determined from (2.21). 
$g_{0}$ is an arbitrary constant for a fixed Riemann surface $\Gamma$.
\par
{\it Remark}. It is possible to choose the local parameters in $\infty $ and
$P_{0}$ so that $g_{1}=g_{2}=0$. If $j = 1/2$, then $g_{1}\equiv 0$ and 
one of the local parameters in $\infty $ or in $P_{0}$ may be chosen in 
an arbitrary way.
\par
The formula (2.22) can be proved from the following connection
between the Baker-Akhiezer forms and the $\tau $-function (for $j=0$ see
\cite{3}):
$$
\psi _{j}(\lambda ,t_{0},\vec{t})=
X_{j}(\lambda )\tau _{j}(t_{0}+1,\vec{t})/\tau _{j}(t_{0},\vec{t}),
$$
$$
\psi ^{*}_{j}(\lambda ,t_{0},\vec{t})=
X^{*}_{j}(\lambda )\tau _{j}(t_{0}-1,\vec{t})/\tau _{j}(t_{0},\vec{t}).
\eqno(2.23)
$$
Relation (2.23) follows from 2.7 and (2.18),(2.19).
\par
\medskip
\noindent {\bf Chapter 3. Virasoro action on the KP theory objects.}
\par
\medskip
3.1. {\em The correspondence between nonlinear equations and variations
of Riemann surfaces}. Now we calculate the action of the vector
fields $v$ on the objects of soliton theory.
\par
{\sl Theorem 3.1}. Let vector field $v$ on $S$ act on the Riemann surface
$\Gamma $, and let the divisor $\gamma _{1},\ldots  ,\gamma _{g}$ and the 
local parameter $z=1/\lambda $ be mapped by E. Then the variation 
of the KP solution $\chi$ is given by:
$$
\partial \chi (t_{0},\vec{t})/\partial \beta  = 
-(2\pi i)^{-1}\oint _{S}(L_{v}\psi _{j}(t_{0},\vec{t}))
\psi ^{*}_{j}(t_{0},\vec{t}) ,
\eqno(3.1)
$$
$$
(\partial _{y}-\partial _{x}^2-2\chi _{x}) \psi _{j}(t_{0},\vec{t}) = 0 ,
\eqno(3.2)
$$
$$
(\partial _{y}+\partial _{x}^2+2\chi _{x}) \psi ^{*}_{j}(t_{0},\vec{t}) = 0 ,
\eqno(3.3)
$$
and the variation of the Baker-Akhiezer function is the following
$$
\partial \psi _{j}(\lambda ,t_{0},\vec{t})/\partial \beta  = 
(2\pi i)^{-1}\oint _{S}(L_{v}\psi _{j}(\mu ,t_{0},\vec{t})) 
\omega _{j}(\lambda ,\mu ,t_{0},\vec{t}),\ \mu \in S , 
\eqno(3.4)
$$
where $L_{v}$ is the Lie derivative.
\par
\noindent The proof follows from (1.12) and asymptotics of $\omega _{j}$
when $\lambda \rightarrow \infty$.
\par
Formula (3.4) solves the Krichever-Novikov problem of calculation of the 
vector field's action on the Baker-Akhiezer function \cite{7}.
\par
{\it Remark}. (3.1-3) is an integrable Lagrangean system which
commutes with the ordinary (commutative) KP hierarchy.  The  $L-A$ pair
for this system is given by (3.2) and (3.4).  Using series like (A.3)
we obtain a more familiar evolution form. This system is discussed in
Appendix A.
\par
{\it Remark}. The variation of the wave function $\psi _{j}$ with the help of
kernel (2.13) may be considered as a correct analytical form of
infinitesimal Zakharov-Shabat dressing which is valid in the
finite-gap case as well as in the decreasing one.
\par
{\em Theorem 3.2}. The commutator of flows (3.1-3) as well as (3.4)
coincides with the commutator of the vector fields $v$ if the following
natural assumptions are valid: 1) The spectral parameter $\lambda =1/z$ and
the divisor $D$ are mapped by $E$ (see 1.4). 2) The asymptotic behaviour
of $\psi _{j}$ and $\psi ^{*}_{j}$ under the action of $v$ remains fixed: 
$\psi _{j}(t_{0}) ~ z^{-to}(dz)^{j}\cdot \exp \sum  z^{-m}t_{m}$, 
$\psi ^{*}_{j}(t_{0}) ~ z^{to-2}(dz)^{1-j}\exp(-\sum  z^{-m}t_{m})$. 
3) We compare Baker-Akhiezer forms on different Riemann surfaces. 
The connection between the points of different Riemann surfaces is 
established by the map E. The Baker-Akhiezer functions are compared in 
these points. 4) In accordance with section 1.4, the vector fields 
$v$ are assumed to be independent of $\beta$ functions of $z$.
\par
\noindent The proof follows from direct calculation.
\par

3.2. {\em Isospectral and non-isospectral symmetries}. Consider the
space $V$ of all vector fields on S. It is known that for $g>1$ $V$ can be
presented as a direct sum $V=V_{+}\oplus V_{0}\oplus V_{-}$. Here $V_{+}$
and $V_{-}$ are the fields which can be analytically continued to 
the regions $\Gamma _{+}$ and $\Gamma _{-}$, 
respectively, and  $\dim V_{0}=3g-3=\dim \hbox{Ker} \bar{\partial }_{2}$.  
The set of Riemann
surfaces of genus $g>1$ can be parameterized by $3g-3$ complex
parameters. This set is called a moduli space.  There are no natural
coordinates on the moduli space but locally we may use $3g-3$
independent elements of Riemann matrix $B_{ij}$. The action of the vector
fields from $V_{0}$ at the moduli space is nondegenerate.  Therefore, the
times of corresponding higher KP equations form local coordinates on
the moduli space.
\par
Symmetries corresponding to  $v\in V_{+},V_{-}$  do not change the Riemann
surface (one can see it from (1.14)), so they are isospectral. Vector
fields $v\in V_{-}$ change the local parameter near $\infty $. Symmetry action
corresponding to $v\in V_{+}$ comes to the ordinary higher KP symmetry
action.
\par
Let us note that for a Riemann surface with two marked points
there exists a natural basis of vector fields corresponding to the
decomposition $V_{+}\oplus V_{0}\oplus V_{-}$- Krichever-Novikov basis
\cite{7}. The action of fields $v\in V_{+}$ on KP theory objects was studied 
in \cite{7}.
\par
3.3. {\em $\tau$-function variation by complex structure. Variation
of $\det\bar\partial_{j}$}.  The action of the vector field $v$ on $\Gamma$ 
results in the transformation of the flag space and $\tau $-function 
(see (2.19)). Using explicit representation 2.5 we obtain
$$
\partial \ln\tau _{j}(t_{0},\vec{t})/\partial \beta  = 
(2\pi i)^{-1}\oint _{S}L_{v}(\lambda )\omega ^{\hbox{reg}}_{j}
(\lambda ,\mu ,t_{0},\vec{t})\mid _{\lambda =\mu },
\eqno(3.5)
$$
where
$$ 
\omega ^{\hbox{reg}}_{j}(\lambda ,\mu ,t_{0},\vec{t}) = 
\omega _{j}(\lambda ,\mu ,t_{0},\vec{t})-
\omega ^{0}_{j}(\lambda ,\mu ,t_{0},\vec{t}).
$$
\par
The "naive" calculation of variation of  $\det\bar{\partial }_{j}$ on 
${\Bbb B}_{j}(t_{0},\vec{t},D)$ by the complex structure gives  
$$
\delta \det\bar{\partial }_{j}= 
\det\bar{\partial }_{j}\cdot 
(\det(1+\bar{\partial }^{-1}_{j}\kappa \partial _{j})-1) = \det\partial_{j}
\cdot \hbox{Tr}\partial _{j}\bar{\partial }^{-1}_{j}\kappa  =
$$
$$ 
=\det\bar{\partial }_{j}\cdot (2\pi i)^{-1}\oint _{S}
L_{v}(\lambda ) \omega ^{\hbox{reg}}_{j}(\lambda ,\mu ,t_{0},\vec{t})
\mid _{\lambda =\mu }
$$. 
We use $\delta \bar{\partial }_{j}=\kappa \partial _{j}$, 
where $\kappa$ is the Beltrami differential corresponding to the
variations of the complex structure. In our case it is a $\delta $-type
function on  S.  So  if we use the same regularization we have
$$
\delta  \ln \det\bar{\partial }_{j}= 
\delta  \ln \tau _{j}(t_{0},\vec{t}).
$$
\par
3.4. {\em Virasoro action on the $\tau$-function. Explicit formulas.} From
(2.20), (3.5) it follows that $\tau $-function obeys differential equations:
$$\partial \tau _{j}(t_{0},\vec{t})/\partial \beta _{m}= 
L^{j}\tau _{j}(t_{0},\vec{t})
\eqno(3.6)
$$
where the time $\beta _{m}$ corresponds to the vector field 
$v=\lambda ^{m+1}d/d\lambda$, 
$$
L^{j}_{m}= \sum  (kt_{k}\partial _{k+m}+
\frac12 (\partial _{k}\partial _{m-k})) + 
(t_{0}-2j+(j-\frac12)(m+1))\partial _{m},\ m > 0 ,
$$
$$
L^{j}_0= \sum  kt_{k}\partial _{k}+ \frac12 (t_{0}-2j)^{2} + (j-1/2)(t_{0}-2j),
\eqno(3.7)
$$
$$
L^{j}_{m}=\sum (kt_{k}\partial _{k+m}+ \frac12 k(m-k)t_{-k}t_{k-m}) - 
m(t_{0}-2j+(j-\frac12)(m+1))t_{-m},\ m < 0.
$$
Here $\partial _{k}=\partial /\partial t_{k}$ and indices are assumed 
to be positive. Operators $L^{j}_{m}$ form the Virasoro algebra with 
central charge  $c_{j}= 6j^{2}-6j+1$.
\par
{\it Remark.} Using the expression for KP solution $\chi (t_{0},\vec{t})
=\partial \ln\tau _{j}(t_{0},\vec{t})/\partial t_{1}$ we obtain 
the formula for the variation of $\chi$ derived in the different way 
described in 3.1.
\par
{\it Remark}. Substituting (2.22) into (3.6) we obtain variations of
$B_{mn}$, $Q_{ik}$, $\vec{U}_{k}$, $\vec{\zeta }$, $h_{k}$, $g_{2}$, 
$g_{1}$, $g_{0}$ and other geometrical objects on $\Gamma $ by
varying the complex structure (see Appendix B).
\par
{\em Appendix A. Higher KP symmetries}. The ordinary higher KP
equations corresponding to the times $t_{m}$ are the symmetries of the KP
itself (i.e., they commute with it). They mutually commute and do not
explicitly depend upon $t_{m}$. These symmetries are a part of a broader
hierarchy parameterized by two integers $m$, $n$ (see \cite{10} and references
therein). These equations explicitly depend upon $t_{m}$ and commute with
the ordinary KP-hierarchy, but in general they do not commute with
each other. They are
$$
\partial \chi /\partial \beta _{mn} =
\hbox{res}\!\!\left.\vphantom{)}\right._{\lambda =\infty }(\lambda ^{m}
((\partial /\partial \lambda )^{n}
w(\lambda ,\vec{t}))w^{*}(\lambda ,\vec{t})),
\eqno(A.1)
$$
where   $w(\lambda ,t)$  and $w^{*}(\lambda ,\vec{t})$  satisfy 
the auxiliary linear problem
$$
(\partial _{y}-\partial ^{2}_{x}-2\chi _{x})w(\lambda ,\vec{t})=0 , \ 
(\partial _{y}+\partial ^{2}_{x}+2\chi _{x})w^{*}(\lambda ,\vec{t})=0
\eqno(A.2)
$$
and have asymptotic behaviour 
$$
\begin{array}{c}
w(\lambda ,\vec{t})=(1+\sum  w_{n}(\vec{t})\lambda ^{-n}))
\exp\sum  t_{n}\lambda ^{n},
\\ \\
w^{*}(\lambda ,\vec{t})=(1+\sum  w^{*}_{n}(\vec{t})\lambda ^{-n})
\exp(-\sum  t_{n}\lambda ^{n}),
\end{array}
\eqno(A.3)
$$
when  $\lambda \rightarrow \infty$; $\beta _{mn}$  is the corresponding time.  
Equations (A1)-(A2) can be
written in the simple Lagrangean form,  $w_{n}, w^{*}_{n}$
and  $\chi $  being independent variables.  Expressing recurrently 
$w_{n}$, $w^{*}_{n}$ via  $\chi $  from
(A.2) and substituting them into (A.1), we obtain a more familiar
form of higher KP equations which are nonlocal evolution equations at
one function $\chi$. For  $n=0$  we have the ordinary (commutative)
KP-hierarchy,  $\beta _{mo}$  being equal to  $t_{m}$.  When  $n=1$  we obtain
conformal symmetries. Tensor properties were not treated in \cite{10}.
\par
Symmetries (A.1) admit another description, one like the
description in \cite{3}.  Let $K = 1 + \sum ^{\infty }_{1}K_{n}\partial^{-n}$, 
where $\partial =\partial /\partial x$, be a
pseudodifferential operator and let   $L = K\circ \partial \circ K^{-1}$, 
$M = K\circ (\sum ^{\infty }_{1}mt_{m}\partial ^{m-1})\circ K^{-1}$, 
$[L,M]=1$, where $\circ $ and $[,]$ denote, respectively, the product and
the commutator in the algebra of the pseudodifferential operators
\cite{26}.  Let $()_-$ be the projector 
$(\sum f_{n}\partial ^{n})_{-}= \sum _{n<0}f_{n}\partial ^{n}$. 
Consider equation
$$
\partial L/\partial \beta _{nm}= [L, (M^{n}\circ L^{m})_{-}] = 0 .
\eqno(A.4)
$$
It is compatible with the ordinary KP hierarchy
$\partial L/\partial t_{k}=[L,(L^{k})_{-}]$.  For $\chi  = K_{1}$ 
one can obtain (A.1).
\par
It was noted \cite{10} that invariant solutions for these symmetries
can be described in terms of the isomonodromy problem \cite{5}, \cite{29}.
\par
{\em Appendix B. The variations of geometrical objects}. From
(2.22) and (3.5) we have for $j=1/2$, $n > 0$:
$$
\partial g_{0}/\partial \beta _{n}= \sum ^{n-1}_{1}Q_{m,n-m}/2 ,
\eqno(B.1)
$$
$$
\partial Q_{mk}/\partial \beta _{n}= m Q{ } _{n+m,k}+ k Q{ } _{m,n+k}+ 
\sum ^{n-1}_{1}Q_{l,m}Q_{n-l,k},
\eqno(B.2)
$$
$$
\partial \vec{\zeta }/\partial \beta _{n}= 0 ,
\eqno(B.3)
$$
$$
\partial \vec{U}_{k}/\partial \beta _{n}= k \vec{U}_{n+k}+ 
\sum ^{n-1}_{1}\vec{U}_{m}Q_{n-m,k},
\eqno(B.4)
$$
$$
\partial \vec{U}_{0}/\partial \beta _{n}= \vec{U}_{n}+ 
\sum ^{n-1}_{1}(Q_{0,n-m}- q_{0,n-m})\vec{U}_{m} ,
\eqno(B.5)
$$
$$
\partial (Q_{0,k}-q_{0,k})/\partial \beta _{n}= 
Q_{kn}+k(Q_{0,k+n}- q_{0,k+n})+ \sum  (Q_{0,l}- q_{0,l})Q_{k,n-l},
\eqno(B.6)
$$
$$
\partial g_{2}/\partial \beta _{n}= 
\sum  (Q_{0,k}-q_{0,k})(Q_{0,n-k}-q_{0,n-k}) + 2(Q_{0,n}-q_{0,n}) ,
\eqno(B.7)
$$
$$
\partial B_{kl}/\partial \beta _{n}= 
2\pi i \sum ^{n-1}_{1}(\vec{U}_{m})_{k}(\vec{U}_{n-m})_{l}.
\eqno(B.8)
$$ 
For $j=1/2$, $n=0$:
$$
\partial Q_{mn}/\partial \beta _{0}= (m+n) Q_{mn},
\eqno(B.9)
$$
$$
\partial \vec{U}_{k}/\partial \beta _{0}= k\vec{U}_{k},
\eqno(B.10)
$$
$$
\partial (Q_{0,k}-q_{0,k})/\partial \beta _{0}= k(Q_{0,k}- q_{0,k}) ,
\eqno(B.11)
$$
$$
\partial g_{2}/\partial \beta _{0}= 1.
\eqno(B.12)
$$
The other derivatives are equal to zero. For $j=1/2$, $n < 0$ we obtain:
$$
\partial Q_{kl}/\partial \beta _{n}= 
kl \delta _{k+n+l,0}+ k \vartheta (k+n)Q{ } _{n+k,l}+ 
l \vartheta (l+n) Q{ } _{k,n+l} 
\eqno(B.13)
$$
$$
\partial \vec{U}_{k}/\partial \beta _{n}= 
k \vec{U}_{n+k}\vartheta (k+n)
\eqno(B.14)
$$
$$
\partial (Q_{0,k}-q_{0,k})/\partial \beta _{n}= 
k\delta _{k+n,0}+ k(Q_{0,k+n}- q_{0,k+n}) \vartheta (k+n)
\eqno(B.15)
$$
The other derivatives are equal to zero. Here $\vartheta (k)=1$ if $n>0$, 
or $=0$ if $n\le 0$.
\par
{\em Appendix C. Krichever-Novikov fermions.} Let $b_{k}$, $c_{k}$ be fermionic
operators with the usual anticommutators: $[b_{n},b_{m}]_{+}= 0$, 
$[c_{n},c_{m}]_{+}= 0$, $[c_{n},b_{m}] = \delta _{nm}$, and 
$\mid 0> ( <0\mid  )$ be right (left) vacuum vectors with
the properties:
\par
\medskip
\noindent 
$$
\begin{array}{rlcrl}
b_{n}\mid 0>=0 & (n\ge 0), &&  c_{n}\mid 0>=0 & (n<0)
\\
<0\mid b_{n}=0 & (n<0), &&  <0\mid c_{n}=0 & (n\ge 0)
\end{array}
\eqno(C.1)
$$
Put $\mid k>=C_{k}\mid 0>$, $<k\mid =<0\mid B_{k}$  where $\mid k>$ 
($<k\mid$ ) denotes states with the ``charge'' $k$ ($-k$) and
\par
$$
B_{k}=\left\{
\begin{array}{lr}
c_{-1} \cdots c_k & (k< 0) \\
1& (k=0) \\
b_{0} \cdots b_{k-1} & (k>0)  
\end{array}\right.,\ \ 
C_{k}=\left\{
\begin{array}{lr}
b_{k} \cdots b_{-1} & (k<0) \\
1 & (k=0) \\
c_{k-1} \cdots  c_{0} & (k>0)
\end{array}\right..\ \ 
\eqno(C.2)
$$
\par
\medskip
Let us introduce the following fermion operators on the Riemann
surface $\Gamma $ by analogy with \cite{8},\cite{9}:
$$
b(\gamma )= \sum  b_{n}\psi _{j}(\gamma ,n,\vec{t}), \ \
c(\gamma )= \sum  c_{n}\psi ^{*}_{j}(\gamma ,n+1,\vec{t}),\ \ 
\gamma \in \Gamma ,
$$
which are $j$- and $1-j$-forms on $\Gamma $ and $n=t_{0}$ (see 2.3). Now all
correlation functions are expressed in terms of Baker-Akhiezer
functions.  From (2.14) it follows:
$$
<t_{0}\mid b(\gamma '{})c(\gamma )\mid t_{0}> = 
\omega _{j}(\gamma '{},\gamma ,t_{0},\vec{t}) ,
\eqno(C.3)
$$
where $\omega _{j}(\gamma '{},\gamma ,t_{0},\vec{t})$ is the 
Cauchy-Baker-Akhiezer kernel (see 2.4).
\par
The Baker-Akhiezer functions can be expressed via fermions in a
way similiar to \cite{3}:
$$
\psi _{j}(\gamma ,n,\vec{t}) = <n\mid b(\gamma ,\vec{t})\mid n+1>, 
\psi ^{*}_{j}(\gamma ,n+1,\vec{t}) = <n+1\mid c(\gamma ,\vec{t})\mid n>. 
\eqno(C.4)
$$
\par
{\bf Acknowledgements}. We are grateful to S.P.Novikov, I.M.Krichever, 
B.A.Dubrovin, S.M.Natanzon, A.M.Levin, A.Yu.Morozov and
A.S.Schwartz for useful discussions. One of the authors (A.O.) is
grateful to V.E.Zakharov for support, E.I.Schulman for interest, and
K.Gotfried and A.Le`Clair for help while he worked at Cornell
University.

\end{document}